\documentclass[a4paper,11pt]{article}
\pdfoutput=1 

\usepackage{jheppub} 

\usepackage[T1]{fontenc} 

\usepackage{tensor}
\usepackage{graphicx,subcaption}
\graphicspath{ {figures/} }	
\usepackage{amsfonts}
\usepackage{xparse}
 \usepackage{amsmath}
 \usepackage{amssymb}
 \usepackage{mathtools}
\usepackage{epsfig}
\usepackage{pdfpages}
\usepackage{bbm}
\usepackage{graphicx,epstopdf}
\usepackage[numbers]{natbib}
\usepackage[makeroom]{cancel}
\usepackage{hyperref}
\usepackage{array}
\usepackage[export]{adjustbox}
\usepackage[normalem]{ulem}
\usepackage{tcolorbox}
\usepackage{romannum}
\usepackage[utf8]{inputenc}

  \def\S{\mathcal{S}}

\title{Bulk-boundary thermodynamic equivalence: a topology viewpoint}

\author[a]{Ming Zhang,}
\author[b]{Jie Jiang \footnote{Corresponding author}}

\affiliation[a]{Department of Physics, Jiangxi Normal University,\\ Nanchang 330022, China}
\affiliation[b]{College of Education for the Future, Beijing Normal University, \\ Zhuhai 519087, China}

\emailAdd{mingzhang@jxnu.edu.cn}
\emailAdd{jiejiang@mail.bnu.edu.cn}

\abstract{Setting the cosmological constant to be dynamical, we study the bulk and boundary thermodynamics of charged Anti-de Sitter black holes. We develop mass/energy formulas in terms of thermodynamic state functions for the extended thermodynamics, mixed thermodynamics, and  boundary conformal field theory thermodynamics. We employ the residue method to study the topological properties of the phase transitions. Our analysis reveals that the bulk and boundary thermodynamics are topologically equivalent for both criticalities and first-order phase transitions in the canonical ensembles, as well as for the Hawking-Page(-like) phase transitions in the grand canonical ensembles. Additionally, those three kinds of phase transitions are shown to be distinguished by their unique topological charges. Our results exemplify the gravity-gauge duality in terms of topology.}

\begin{document}

\maketitle
\flushbottom

\section{Introduction}

Black hole thermodynamics has attracted growing attention ever since the discovery that the mass variation of a black hole resembles that of ordinary thermodynamic systems \cite{Bardeen:1973gs}. In the semiclassical setting \cite{Bekenstein:1973ur,Hawking:1975vcx}, the well-known relations between temperature $T$, Bekenstein-Hawking entropy $S$, surface gravity $\kappa$, and horizon area $A$ as given by 
\begin{equation}\label{temka}
T = \frac{\kappa}{2 \pi}, \quad S = \frac{A}{4 G_{N}},
\end{equation}
where $G_N$ is Newton's constant, has triggered an abundance of explorations into the thermodynamic properties of the black hole system. The thermodynamics of asymptotically Anti-de Sitter (AdS) black holes has received special attention, mainly due to their dual interpretation in terms of thermal states in the dual conformal field theory (CFT) via the AdS/CFT correspondence \cite{Maldacena:1997re}. One famous example of this is the Hawking-Page transition between a large black hole and thermal radiation in the bulk AdS spacetime and its complementary thermalization transition in the boundary strongly coupled dual CFT \cite{Hawking:1982dh}. In this way, some indiscernible phases in the CFTs can be probed \cite{Cong:2021jgb}, such as the triple points in the QCD diagram \cite{Cui:2021qpu}.

In $D$-dimensional spacetime, due to viewing the negative cosmological constant $\Lambda$ as a thermodynamic state variable \cite{Kastor:2009wy,Dolan:2010ha,Dolan:2011xt,Cvetic:2010jb}, i.e.
\begin{equation}\label{plam}
P=-\frac{\Lambda}{8 \pi G_N}, \quad \Lambda=-\frac{(D-1)(D-2)}{2 L^{2}},
\end{equation}
where $L$ is the bulk curvature radius, investigations concerning the phase behavior of asymptotically AdS black holes have uncovered a wide landscape. This investigation has included Van der Waals-like phase transition \cite{Chamblin:1999tk,Chamblin:1999hg,Cvetic:1999ne,Kubiznak:2012wp}, superfluid behaviors \cite{Hennigar:2016xwd}, black hole microstructures \cite{Wei:2019uqg}, and multicritical phase transitions \cite{Tavakoli:2022kmo,Wu:2022plw}. Higher-dimensional origin of the extended black hole thermodynamics was proposed recently \cite{Frassino:2022zaz}.  For more details, refer to \cite{Kubiznak:2016qmn} and the references therein and thereof. The first law of the {\it extended thermodynamics} for black holes and its dimension-dependent integral Smarr relation read
\begin{align}
d M&=T dS+\Phi dQ+V dP+\Omega_b dJ, \label{bulkfi}\\
M&=\frac{D-2}{D-3}(T S+\Omega_b J)+\Phi Q-\frac{2}{D-3} P V,\label{bulksm}
\end{align}
where the mass $M$ is interpreted as thermodynamic enthalpy instead of internal energy, $Q$ the electric charge, and $J$ the angular momentum. $\Phi$ and $\Omega_b$ are conjugate potentials, and $V$ is the thermodynamic volume.

Recently, there has been a suggestion in \cite{Visser:2021eqk,Cong:2021fnf,Cong:2021jgb} that  the Newton's constant $G_N$ should be considered as a thermodynamic variable in its own right. This idea originated from the holographic interpretation of black hole chemistry \cite{Kastor:2014dra,Dolan:2014cja,Zhang:2014uoa}, which claims that the bulk thermodynamics of black holes is equivalent to the boundary thermodynamics of strongly coupled gauge theories in the large-$N$ limit. Specifically, we have the equivalence of the bulk AdS and boundary CFT partition functions: $Z_{\mathrm{CFT}}=Z_{\mathrm{AdS}}$ \cite{Gubser:1998bc,Witten:1998qj}. Therefore, according to the holographic duality identification \cite{Henningson:1998gx,Myers:2010tj,Karch:2015rpa}, for the Einstein gravity, we have
\begin{equation}\label{hdc}
C=\frac{\Omega_{D-2} L^{D-2}}{16\pi G_N},
\end{equation}
where $\Omega_{D-2}$ is the volume of a unit $D-2$-sphere, $C$ can be viewed as the central charge of the dual conformally symmetric $SU(N)$ gauge theory at large $N$ which counts the number of field degrees of freedom \cite{Gadde:2023daq}. Thus, varying the bulk cosmological constant means that both the central charge $C$ of the CFT (or the number of colors $N$) and the boundary volume $\mathcal{V}$ change \cite{Karch:2015rpa}. If we propose a theory that the Newton's constant $G_N$ is designed to be dynamic (which from the perspective of quantum field theory is possible), then possibly the central charge may be kept fixed so that the dual CFT can remain unchanged even if $\Lambda$ is rescaled. In this paradigm, the first law of bulk black hole spacetime thermodynamics should be written in a mixed form \cite{Cong:2021fnf}
\begin{equation}
d M=TdS+\phi dQ+V_{C}dP+\mu_b dC+\Omega_b dJ,\label{154807}
\end{equation}
where $\phi, V_C, \mu_b$ are the redefined electric potential, thermodynamic volume, and a new conjugate thermodynamic potential (also named "color susceptibility"). With this first law of the {\it mixed thermodynamics}, one can study the bulk thermodynamics with an invariable boundary central charge.

The bulk black hole thermodynamics is well developed along the aforementioned line. On the other hand, we have the {\it CFT thermodynamics} first law or the thermodynamic fundamental equation for the thermal states in the boundary CFT, which reads \cite{Visser:2021eqk}
\begin{equation}\label{hflb}
d E =\mathcal{T} d\mathcal{S}-pd\mathcal{V}+\varphi d\mathcal{Q}+\mu dC+\Omega d\mathcal{J},
\end{equation}
where $E$ is the energy (not enthalpy), and $p,\mathcal{V}$ are the CFT pressure and volume (which are not dual to the bulk pressure $P$ and volume $V$, respectively \cite{Johnson:2014yja}). $\mathcal{T}, \mathcal{S}, \varphi, \mathcal{Q}, \Omega, \mathcal{J}, \mu$ are boundary quantities corresponding to their bulk counterparts $T, S, \Phi, Q, \Omega_b, J, \mu_b$, respectively. The variation of $C$ corresponds to a change of the field theory.

In the following, we will first provide compact generalized mass/energy formulas for the charged AdS black holes in the extended thermodynamics, mixed thermodynamics, and CFT thermodynamics in section \ref{sec2}. Then, in section \ref{sec3}, we will propose a residue method for studying the topologies of phase transitions in the extended thermodynamics, mixed thermodynamics, and CFT thermodynamics for the charged AdS black holes in both the canonical and grand canonical ensembles. Specifically, we will respectively investigate the criticalities and first-order phase transitions in the canonical ensemble and the Hawking-Page(-like) phase transitions in the grand canonical ensemble for the bulk and boundary of the charged AdS black hole spacetime. Conclusion and discussion will be given in the last section. We aim to answer the question of whether there is any indication of bulk-boundary thermodynamic equivalence from the viewpoint of topology. For concreteness, we consider the four-dimensional ($D=4$) charged AdS black hole, and related extensions are straightforward.

\section{Generalized mass/energy formulas of the charged AdS black holes \label{sec2}}

\subsection{Extended thermodynamics}
Thermal states in the boundary CFT correspond to the classical black hole solutions. We here consider the  Einstein-Maxwell theory, whose action is
\begin{equation}\label{actb}
I=\frac{1}{16 \pi G_N} \int d^{4} x \sqrt{-g}\left(R-2 \Lambda-F^2\right).
\end{equation}
The least action principle yields a charged spherically symmetric Reissner-Nordström-AdS (RN-AdS) spacetime and its corresponding gauge field as
\begin{align}
d s^2&=-f(r) d t^2+\frac{d r^2}{f(r)}+r^2 d \Omega_{2}^2,\\
A&=\left(- \frac{G_N Q}{r}+\Phi\right) d t,\\
f(r)&=1+\frac{r^2}{L^2}-\frac{2G_NM}{r}+\frac{G_N^2 Q^2}{r^{2 }},
\end{align}
where $d \Omega_{2}^2$ is the metric of a unit 2-sphere, and $M, Q$ are the ADM mass and physical electric charge of the black hole. The event horizon $r_h$ of the black hole is the larger real root of $f(r)=0$. By selecting the gauge $A_t\left(r_h\right)=0$, we have 
\begin{equation}
\Phi= \frac{G_NQ}{r_h}.
\end{equation}
The bulk  extended thermodynamics first law and Smarr formula for the black hole, which includes variations of the parameters $G_N$ and $\Lambda$, are given by \cite{Cong:2021jgb}
\begin{align}
M&=\frac{\kappa A}{4 \pi G_N}+\Phi Q+\frac{V \Lambda}{4 \pi G_N},\\
d M&=\frac{\kappa}{8 \pi G_N} d A+\Phi d Q-\frac{V}{8 \pi G_N} d \Lambda-(M-\Phi Q) \frac{d G_N}{G_N},\label{bulkF}
\end{align}
where 
\begin{align}
\kappa=\frac{1}{2 r_h}\left(1+\frac{3r_h^2}{L^2}-\frac{G_N^2 Q^2}{r_h^{2 }}\right), A=4\pi r_h^{2},
V=\frac{4\pi}{3} r_h^3.
\end{align}

We propose an alternative mass formula that complements the Smarr mass formula, which is given by
\begin{equation}\label{mass1}
M=-\frac{A^{3/2} \Lambda }{48 \pi ^{3/2} G_N}+\frac{\sqrt{\pi } G_N Q^2}{\sqrt{A}}+\frac{\sqrt{A}}{4 \sqrt{\pi } G_N},
\end{equation}
such that
\begin{equation}\label{tempet}
\begin{aligned}
\frac{\kappa}{8 \pi G_N}=\left(\frac{\partial M}{\partial A}\right)_{Q, \Lambda, G_N}, \quad\Phi=\left(\frac{\partial M}{\partial Q}\right)_{A, \Lambda, G_N},
\end{aligned}
\end{equation}
\begin{equation}
\begin{aligned}
\frac{-V}{8 \pi G_N}=\left(\frac{\partial M}{\partial \Lambda}\right)_{A, Q, G_N},\quad \frac{\Phi Q-M}{G_N}=\left(\frac{\partial M}{\partial G_N}\right)_{A, Q, \Lambda}.
\end{aligned}
\end{equation}
The mass formula is a $G_N$-including version of the one proposed in \cite{Caldarelli:1999xj}. It can also be obtained from dimensional analysis. The temperature $T$, entropy $S$ of the black hole in the extended thermodynamics can be obtained by applying (\ref{temka}).

\subsection{Mixed thermodynamics}
For the considered Einstein gravity, the AdS/CFT holographic dictionary for the CFT central charge  in terms of the bulk AdS radius, Newton's constant is given by (\ref{hdc}). After incorporating this central charge into the bulk first law (\ref{bulkF}), we have the first law of the mixed thermodynamics as
\begin{equation}
d M=T d S+\phi d Q+V_C d P+\mu_b d C,
\end{equation}
where the newly defined thermodynamic volume and the chemical potential read
\begin{equation}
V_C=\frac{2 M}{8 P}-\frac{\sqrt{\pi  G_N} Q^2}{4 P \sqrt{S}}, \mu_b=\frac{2 P (V_C-V)}{2 C}.
\end{equation}
We see that this mixed form of thermodynamics allows us to study the bulk thermodynamics in the context of fixed boundary central charge. Note that the redefined volume $V_C$ here is a bit different from the one given in \cite{Cong:2021fnf} since there is a different convention for the electromagnetic term in the bulk action (\ref{actb}).
We can also construct a mass formula encoding the central charge as
\begin{equation}\label{mass2}
\begin{aligned}
M^2=&\frac{\sqrt{P} S^3}{8 \sqrt{6} \pi ^{5/2} C^{3/2}}+\frac{\sqrt{3 \pi } Q^4}{16 \sqrt{2CP}  S}+\frac{\sqrt{P} S^2}{\sqrt{6} \pi ^{3/2} \sqrt{C}}\\&+\sqrt{\frac{2}{3 \pi }} \sqrt{CP} S+\frac{Q^2 S}{8 \pi  C}+\frac{Q^2}{2}.
\end{aligned}
\end{equation}
Then  $T, \phi, V_C, \mu_b$ can be obtained  via conducting derivatives with respect to their conjugate thermodynamic quantities as
\begin{align}\label{mixedqphi}
T&=\left(\frac{\partial M}{\partial S}\right)_{Q, P, C},\, \phi=\left(\frac{\partial M}{\partial Q}\right)_{S, P, C},\\
V_C&=\left(\frac{\partial M}{\partial P}\right)_{Q, S, C},\, \mu_b=\left(\frac{\partial M}{\partial C}\right)_{S, Q, P}.
\end{align}
We do not list the specific forms of these quantities for brevity. It is remarkable that this mass formula makes us convenient to calculate quantities in the mixed thermodynamics, which employs the boundary central charge to prescribe the bulk thermodynamics.

\subsection{CFT thermodynamics}
To construct the boundary CFT thermodynamics of the charged AdS black hole, we need to use more holographic relations between  bulk and boundary quantities. We here set the curvature radius of the boundary to be $R$, which is different from the bulk AdS radius $L$.  The metric of the CFT admitting conformal scaling invariance reads \cite{Gubser:1998bc,Witten:1998qj,Ahmed:2023snm}
\begin{equation}
d s^2=\omega^2 \left(-d t^2+L^2 d \Omega_{2}^2\right).
\end{equation}
We can choose the conformal gauge as $\omega=R/L$ \cite{Visser:2021eqk,Cong:2021jgb} (Other selection of the gauge does not change our result). Thus the spatial volume of the boundary sphere can be expressed in terms of boundary curvature radius as
\begin{equation}
\mathcal{V}=\Omega_{2} R^{2}.
\end{equation}
We can have the holographic dictionary between bulk quantities $M, \kappa, A, \Phi, Q$ and their boundary CFT counterparts $E, \mathcal{T}, \mathcal{S}, \varphi, \mathcal{Q}$ as \cite{Chamblin:1999tk,Savonije:2001nd,Visser:2021eqk}
\begin{align}
E&=M \frac{L}{R}, \mathcal{T}=\frac{\kappa}{2 \pi} \frac{L}{R}, \mathcal{S}=S=\frac{A}{4 G_N},\\
\varphi&=\frac{\Phi}{L} \frac{L}{R}, \mathcal{Q}=Q L.
\end{align}
Note that the boundary entropy is related to the bulk horizon area. The $L/R$ term arises as the difference between the bulk time and the boundary time.

The first law and the dimension-independent Euler relation for the boundary CFT dual to their bulk AdS counterparts (\ref{bulkfi}) and (\ref{bulksm}) read \cite{Karch:2015rpa,Visser:2021eqk}
\begin{align}
d E&=\mathcal{T} d \mathcal{S}+\varphi d \mathcal{Q}-p d \mathcal{V}+\mu d C,\\
E&=\mathcal{T} \mathcal{S}+\varphi \mathcal{Q}+\mu C.
\end{align}
The Euler relation can be derived from the scaling properties of finite temperature gauge theories in the large-$N$ 't Hooft limit \cite{tHooft:1973alw}. The pressure and chemical potential are constrained to be
\begin{align}
p&= \frac{E}{2\mathcal{V}},\\
\mu&=\frac{E-\mathcal{T} \mathcal{S}-\varphi \mathcal{Q}}{C}.
\end{align}
The former one is the equation of state for the CFT and the latter one arises as the proportionality between $C$ and $E, \S, \mathcal{Q}$ \cite{Ahmed:2023snm}.

We find that the mass formula (which can also be named the internal energy formula) of the boundary field theory can be written in terms of the boundary thermodynamic  quantities as
\begin{equation}\label{mass3}
E=\frac{4 \pi  C \mathcal{S}+\pi ^2 \mathcal{Q}^2+\mathcal{S}^2}{2 \pi \sqrt{C \mathcal{S} \mathcal{V}}}.
\end{equation}
Then it is straightforward to show that
\begin{equation}\label{tempcf}
\mathcal{T}=\left(\frac{\partial E}{\partial \mathcal{S}}\right)_{\mathcal{Q}, \mathcal{V}, C},\, \varphi=\left(\frac{\partial E}{\partial \mathcal{Q}}\right)_{\mathcal{S}, \mathcal{V}, C},
\end{equation}
\begin{equation}
p=-\left(\frac{\partial E}{\partial \mathcal{V}}\right)_{\mathcal{S}, \mathcal{Q}, C},\, \mu=\left(\frac{\partial E}{\partial C}\right)_{\mathcal{S}, \mathcal{Q}, \mathcal{V}}.
\end{equation}
We should point out that the energy formula (\ref{mass3}), as well as the mass formulas in (\ref{mass1}) and (\ref{mass2}),  not only makes us convenient to calculate the full thermodynamic quantities in the canonical ensemble but also makes us capable of having thermodynamic quantities in other ensembles. For instance, if we are to consider one other ensemble in the CFT thermodynamics, say, the one with fixed $C, \mathcal{Q}, p$, we just need to calculate the free energy $F_p=E-\mathcal{T} \S+p\mathcal{V}$, and then use the differential relation $dF_p=-\S d\mathcal{T}+\mathcal{V}dp+\mu dC$.

\section{Topologies of the phase transitions}\label{sec3}

\subsection{Residue method}
To the end of investigating the topologies of the phase transition, we not only need mass formulas to conveniently calculate the wanted thermodynamic quantities but also have to choose a method to calculate the topologies. Recently, a topological current \cite{Duan:1979ucg} method  was proposed for calculating the topology of the criticality and Hawking-Page phase transition of the bulk thermodynamics of black holes \cite{Wei:2021vdx,Yerra:2022coh,Bai:2022klw}. In this method, one can construct a (somehow arbitrary) two-dimensional vector field and study its winding number around defect points in some two-dimensional plane where the vector field lies. Here we introduce a different way to conduct the topology calculation. This way, we name it the {\it residue method}, typically depends on the residue calculation in the complex analysis. We will now  consider a single-valued function $\mathcal{F}: \mathbb{C}\rightarrow \mathbb{C}$ with a single complex variable $z$. We suppose that the complex function is holomorphic on the complex plane except at some isolated pole points $z_1, z_2, \ldots, z_n$.
We can endow topological numbers (or in other words, topological charges) onto some specific isolated points, say, $z_k$, which can be either real or complex and  it can be a pole point of order $n$. The topology charge $w_k$ of $\mathcal{F}(z)$ at the specific point can be defined as
\begin{equation}\label{omek}
w_k\equiv \text{Sgn}\left[ \frac{1}{2\pi i}\oint_{\gamma}  \mathcal{F}(z)dz\right]=\text{Sgn}[\text{Res} \mathcal{F} (z_k)],
\end{equation}
where $\gamma$ is the counter-clockwise oriented boundary of a small region around the pole point $z_k$,  $\text{Sgn}$ represents the sign function, and $\text{Res}$ denotes the residue. The local topological property of the characteristic function $\mathcal{F}(z)$ can be reflected by the local topological charge at its $n$-order pole point.

In what follows, we will investigate the phase transition of the charged AdS black hole in the canonical and grand canonical ensembles. In the former ensemble, we can find the first-order phase transition and criticality phenomenon while the Hawking-Page phase transition (which is also first-order) can be found in the latter ensemble. To study the topologies of the criticalities and the first-order phase transitions in the canonical ensemble as well as the Hawking-Page(-like) phase transitions in the grand canonical ensemble, we define a universal characteristic function
\begin{equation}\label{chf}
\mathcal{F}(z)=\frac{1}{\partial_{zz}\mathbb{T}(z)},
\end{equation}
where $\mathbb{T}$ is the thermodynamic temperature function of the bulk or the boundary system, and we will choose the entropy of the bulk or the boundary as the complex variable $z$. In the two different ensembles, the temperature functions have different forms. In fact, as our object is to show the bulk-boundary thermodynamic equivalence from a viewpoint of topology, a different construction of the characteristic function $\mathcal{F}$ does not matter. We note that recently we have used the residue method to revisit the topologies of the black hole solutions \cite{Fang:2022rsb} as thermodynamic defects \cite{Wei:2022dzw} and kinematic topologies of black holes were firstly studied in \cite{Cunha:2020azh}.

\subsection{Topologies of criticality and first-order phase transition}

In our study of the charged AdS black holes, we aim to investigate their topological properties during phase transitions. Before diving into our analysis, it is worth reviewing some prior research on black hole phase transitions. Specifically, in the context of extended thermodynamics, previous studies such as \cite{Peca:1998cs,Wei:2020kra} have demonstrated that the charged AdS black holes exhibit Hawking-Page phase transitions in the grand canonical ensemble, along with criticality and first-order phase transitions in the canonical ensemble as shown in \cite{Chamblin:1999tk,Chamblin:1999hg,Cvetic:1999ne,Wu:2000id,Kubiznak:2012wp}.
Furthermore, the mixed thermodynamics of charged AdS black holes demonstrates central charge criticality, which was established in \cite{Cong:2021fnf}. Furthermore, in the CFT thermodynamics, criticality and (de)confinement phase transitions were analyzed in \cite{Cong:2021jgb}. Keeping these prior works in mind will help contextualize and inform our investigation of the topological properties of phase transitions in charged AdS black holes.

\subsubsection{Topology of criticality}
We will now study the topologies of the criticalities for the charged AdS black hole in the canonical ensemble, where $(Q, \Lambda)$ are fixed for the extended thermodynamics,  $(Q, C, P)$ are fixed for the mixed thermodynamics, and $(\mathcal{Q}, C, \mathcal{V})$ are fixed for the CFT thermodynamics. In the extended thermodynamics, the temperature can be obtained via (\ref{temka}) and (\ref{tempet}). We now denote the temperature in the extended thermodynamics, mixed thermodynamics, and CFT thermodynamics as $T_E, T_M, \mathcal{T}_C$, respectively, which, according to (\ref{tempet}), (\ref{mixedqphi}), and (\ref{tempcf}), are explicitly given by
\begin{align}
T_E&=\frac{G_N \left(\pi  S-G_N \left(\pi ^2 Q^2+\Lambda  S^2\right)\right)}{4 \pi ^{3/2} (G_N S)^{3/2}},\label{te}\\
T_M&=\frac{1}{2M}\left(\frac{\sqrt{P} \left(16 \pi ^2 C^2+16 \pi  C S+3 S^2\right)}{8 \sqrt{6} \pi ^{5/2} C^{3/2}}-\frac{Q^2 \left(\sqrt{6} \pi ^{3/2} \sqrt{C} Q^2-4 \sqrt{P} S^2\right)}{32 \pi  C \sqrt{P} S^2}\right),\label{tm}\\
\mathcal{T}_C&=\frac{C \mathcal{V} \left(4 \pi  C \S-\pi ^2 \mathcal{Q}^2+3 \S^2\right)}{4 \pi  (C \S \mathcal{V})^{3/2}}.\label{tc}
\end{align}
Then through the relations
\begin{align}
\left(\frac{\partial T_E}{\partial S}\right)_{Q, \Lambda}&=0=\left(\frac{\partial^2 T_E}{\partial S^2}\right)_{Q, \Lambda},\label{canr1}\\
\left(\frac{\partial T_M}{\partial S}\right)_{Q, C, P}&=0=\left(\frac{\partial^2 T_M}{\partial S^2}\right)_{Q, C, P}\label{canr2},\\
\left(\frac{\partial \mathcal{T}_C}{\partial \mathcal{S}}\right)_{\mathcal{Q}, C, \mathcal{V}}&=0=\left(\frac{\partial^2 \mathcal{T}_C}{\partial \mathcal{S}^2}\right)_{\mathcal{Q}, C, \mathcal{V}}\label{canr3},\\
\end{align}
we have the critical points for the three cases as
\begin{align}
(S_{Ec}, \Lambda_{Ec})&=(6 \pi  G_N Q^2, -\frac{1}{12 G_N^2 Q^2}),\\
(C_{Mc}, P_{Mc},  S_{Mc})&=(9 G_N Q^2, \frac{1}{96 \pi  G_N^3 Q^2}, 6 \pi  G_N Q^2),\\
(C_{Cc}, \mathcal{S}_{Cc})&=(\frac{3 \mathcal{Q}}{2}, \pi  \mathcal{Q}).
\end{align}
After individually  setting $z=(S_{Ec}, 0), (S_{Mc}, 0), (\mathcal{S}_{Cc}, 0)$ in (\ref{chf}), we obtain the topological charges
\begin{equation}\label{equa1}
\omega_E=\omega_M=\omega_C=1
\end{equation}
for each critical point by (\ref{omek}). This equality of the topological charge means that there is a topological equivalence between the criticality of the bulk and boundary thermodynamics for the charged AdS black hole in the canonical ensemble.

\subsubsection{Topology of first-order phase transition}

In the canonical ensemble, the first-order phase transition of the charged AdS black hole takes place on the coexistence line, where the temperature and the Helmholtz free energy of  two phases (which are small/large black holes in the bulk and low/high entropy states in the CFT) are equal. The temperature of the charged AdS black holes in the extended thermodynamics, mixed thermodynamics, and CFT thermodynamics  are given by \eqref{canr1}, \eqref{canr2}, and \eqref{canr3}, respectively. The Helmholtz free energy in these systems are individually given by 
\begin{align}
F_E&=M-T_E S=\frac{9 \pi ^2 Q^2 G_N+\Lambda  S^2 G_N+3 \pi  S}{12 \pi ^{3/2} \sqrt{S G_N}},\\
F_M&=M-T_M S,\label{helfm}\\
F_C&=E-\mathcal{T}_C \mathcal{S}=\frac{4 \pi  C \mathcal{S}+3 \pi ^2 \mathcal{Q}^2-\mathcal{S}^2}{4 \pi  \sqrt{C \mathcal{S} \mathcal{V}}},
\end{align}
where we do not show the explicit expression of $F_M$ for simplicity and the mass and temperature  in \eqref{helfm} can be found in \eqref{mass2} and \eqref{tm}.

In the extended thermodynamics, the coexistence line can be determined by $T_E(S_1^E)=T_E(S_2^E)$ and $F_E(S_1^E)=F_E(S_2^E)$. This yields the following expression \cite{Kubiznak:2016qmn}:
\begin{align}S_{1, 2}^E=\frac{\pi\left(\pm\sqrt{9-12 Q G_N \left(3 \Lambda  Q G_N+2 \sqrt{-3 \Lambda }\right)}+4 \sqrt{-3 \Lambda } Q G_N-3\right)}{2 \Lambda G_N},
\end{align}
where $\Lambda\in [\Lambda_{Ec}, 0[$, $S_1^E\in ]\pi G_N Q^2, S_{Ec}]$, and $S_2^E\in [S_{Ec}, \infty[$. It is worth noting that at the point $T_E=T_E[S_1^E (\Lambda=\Lambda_{Ec})]$, the second-order criticality phase transition takes place. Using these calculations, we can obtain the formula for the coexistence line in extended thermodynamics $T_E(S_1^E)$ (or alternatively, $T_E(S_2^E)$). Similarly, we can obtain the entropies $S_{1, 2}^M$ and $S_{1, 2}^{C}$ corresponding to the points where the first-order phase transitions take place in the mixed thermodynamics and CFT thermodynamics, respectively. For the mixed thermodynamics, the entropies are given by
\begin{equation}\label{sm318}
S_{1, 2}^M=2 \pi C-\left(\frac{24 \pi^3 C}{P}\right)^{1/4} Q\mp 2 \pi C \sqrt{1-\left(\frac{3}{\pi}\right)^{1/4} \frac{Q}{P}\left(\frac{2 P}{C}\right)^{3/4}+\frac{3 Q^2}{8} \sqrt{\frac{6}{\pi C^3}}},
\end{equation}
where $C\in [C_{Mc}, \infty[$, and $P=3/32\pi G_N^2 C$ (which can be obtained from \eqref{plam} and \eqref{hdc}). For the CFT thermodynamics, the entropies are given by
\begin{equation}
\label{sc319}\mathcal{S}_{1, 2}^C=\frac{1}{3} \left(2 \pi C\mp\pi  \sqrt{4 C^2-9 \mathcal{Q}^2}\right),\end{equation}
where $C\in [C_{Cc},\infty[$. The coexistence lines for the mixed and CFT thermodynamics are $T_M(S_1^M)$ and $\mathcal{T}_C(\mathcal{S}_1^M)$, respectively. Note that the criticality phase transitions take place exactly at $T_M=T_M[S_1^M (C=C_{Mc})]$ and $\mathcal{T}_C=\mathcal{T}_C[\mathcal{S}_1^C(C=C_{Cc})]$.

To assign topological charges to the coexistence lines of charged AdS black holes in extended thermodynamics, mixed thermodynamics, and CFT thermodynamics, we use the characteristic function \eqref{chf} with $\mathbb{T}$ specified as $T_E(S_1^E)$, $T_M(S_1^M)$, and $\mathcal{T}_C(\mathcal{S}_1^M)$, respectively. This yields
\begin{equation}\label{equax}
\omega_E = \omega_M = \omega_C = 0
\end{equation}
for all points on the coexistence lines where the first-order phase transitions take place. Mathematically, these zero topological charges can be understood as all first-order phase transition points on the coexistence lines are not pole points of the characteristic function. This suggests that topological equivalence exists between bulk and boundary first-order phase transitions. Interestingly, comparing \eqref{equa1} with \eqref{equax}, we can see that there is a sudden change in topological charge whenever the order of the phase transition of charged AdS black holes changes in extended thermodynamics, mixed thermodynamics, or CFT thermodynamics. For a comparable investigation, we refer the readers to \cite{Fan:2022bsq}, which examined the bulk first-order phase transition of the charged AdS black hole by analyzing the winding number interchange of the topological defects in the landscape of the off-shell internal energy.

\subsection{Topology of Hawking-Page(-like) phase transition}
In the grand canonical ensemble, there are no criticalities for the charged AdS black hole  in all the extended, mixed, and CFT thermodynamics. However, there are Hawking-Page phase transitions in the bulk and (de)confinement phase transitions in the boundary CFT, which is Hawking-Page-like \cite{Cong:2021jgb}. Is there still a topological equivalence between  the bulk and boundary thermodynamics? Here we will investigate this issue, still by the characteristic function (\ref{chf}).  In the extended thermodynamics with fixed $(\Phi, \Lambda)$, mixed thermodynamics with fixed $(\phi, C, P)$, and CFT thermodynamics with fixed $(\varphi, C, \mathcal{V})$, the Gibbs free energy functions read
\begin{align}
G_E&=M-T S-\Phi Q= \frac{S \left(G_N \Lambda  S-3 \pi  \left(\Phi ^2-1\right)\right)}{12 \pi ^{3/2} \sqrt{G_N S}}\label{ginbu},\\
G_M&=M-T S-\phi Q,\\
G_C&=E-\mathcal{T} \mathcal{S}-\varphi \mathcal{Q} =-\frac{\mathcal{S} \left(C \mathcal{V} \varphi^2-4 \pi  C+\mathcal{S}\right)}{4 \pi  \sqrt{C \mathcal{S} \mathcal{V}}},
\end{align}
where we do not   show explicitly the thermodynamic potential $G_M$ as it is lengthy. The Hawking-Page(-like) phase transition occurs at the point where the Gibbs free energy vanishes,
\begin{equation}
G_X=0,
\end{equation}
where $X=\{E, M, C\}$. We then get the phase transition points as
\begin{align}
S_{Ec}&=\frac{\pi  \left(\Phi ^2-1\right)}{G_N \Lambda }\label{gcese},\\
S_{Mc}&=\frac{1}{6} \left(\sqrt{2} \sqrt{8 \pi ^2 C^2-\frac{3 \sqrt{6} \pi ^{3/2} \sqrt{C} Q^2}{\sqrt{P}}}+4 \pi  C\right),\label{smc}\\
\mathcal{S}_{Cc}&=\frac{1}{3} C \left(4 \pi -\mathcal{V} \varphi ^2\right).
\end{align}
Here in (\ref{smc}), $Q$ should be replaced by $\phi$ via the relation (\ref{mixedqphi}). 

Besides, we should have the temperature in the grand canonical ensemble for the three kinds of thermodynamics, which can be obtained by reexpressing the temperature (\ref{te}), (\ref{tm}), and (\ref{tc}) in terms of the electric potential instead of the electric charge by employing (\ref{tempet}), (\ref{mixedqphi}), and (\ref{tempcf}), respectively. Then the characterized function (\ref{chf}) can be calculated through the temperature for the grand canonical ensemble.  After denoting $z_c=(S_{Ec}, 0), (S_{Mc}, 0), (S_{Cc}, 0)$ and applying (\ref{omek}),  we have the topological charges
\begin{equation}\label{equa2}
\omega_E=\omega_M=\omega_C=-1
\end{equation}
for the Hawking-Page(-like) phase transition points in the grand canonical ensembles for the extended, mixed, and CFT thermodynamics. This topological equality again shows the definite equivalence between the bulk and boundary thermodynamics for the charged AdS black hole.

\section{Conclusion and discussion}\label{sec8}

We studied the bulk AdS and boundary CFT thermodynamics for the charged AdS black hole by applying the topological method. To elucidate the thermodynamics of the black hole, we obtained the mass/energy formulas (\ref{mass1}), (\ref{mass2}), and (\ref{mass3}) for the extended, mixed, and CFT thermodynamics, expressed in terms of thermodynamic state functions. After introducing the residue method to assign the topology charges to the phase transitions, we investigated the topologies of the black hole in the canonical and grand canonical ensembles. We revealed the topological equivalence between the thermodynamics of the bulk AdS and the boundary CFT through the topological equalities (\ref{equa1}), \eqref{equax}, and (\ref{equa2}) for the criticalities and first-order phase transitions in the canonical ensembles and the Hawking-Page(-like) phase transitions in the grand canonical ensembles, respectively. It is noteworthy that the criticality, first-order phase transition, and the Hawking-Page(-like) phase transition display dissimilar topological charges. We note that the topological charge of the Hawking-Page(-like) phase transition is $-1$, as stipulated by our definition of the characteristic function (\ref{chf}). We have verified that for modified gravity theories like the AdS black hole in Gauss-Bonnet gravity \cite{Cai:2001dz}, the topological charge is 0 for the Hawking-Page phase transition in the extended thermodynamics using the characteristic function (\ref{chf}). The topological charge $-1$ for the charged AdS black hole is quite subtle under the adoption of the characteristic function (\ref{chf}), as the characteristic function $\mathcal{F}$ and the Gibbs function in the grand canonical ensemble share the same factor, and this mutual factor vanishes in a condition of introducing additional parameters, such as the coupling parameter $\alpha$ in the Gauss-Bonnet-AdS black hole. Our study may inspire further research on bulk-boundary systems with angular momentum \cite{Gong:2023ywu,Ahmed:2023dnh}, multicritical phase transitions \cite{Tavakoli:2022kmo,Wu:2022plw}, string theory corrections \cite{Dutta:2022wbh,Sinamuli:2017rhp,Gnecchi:2013mja}, and other topics that complement the investigation of bulk-boundary thermodynamical equivalence \cite{Gibbons:2005vp,Cvetic:2021vxa,Ezroura:2021vrt}.

We would like to clarify two issues here. First, the characteristic function chosen to assign topologies to the phase transitions is not unique. As a principle, a well-defined characteristic function should be able to encode the phase transition information of the thermodynamic systems as topological defects, and the phase transition points should be the pole points of the function. In the canonical ensemble, we choose the characteristic function \eqref{chf} so that the critical points, where the second-order phase transitions take place, are pole points for which topological charges can be assigned. The topological charges in both the extended thermodynamics, mixed thermodynamics, and CFT thermodynamics are $\omega=1$. On the one hand, this signifies a topological equivalence between bulk and boundary thermodynamics (phase transitions). On the other hand, referring to the observation in \cite{Wei:2021vdx} for the topological charge of the criticality in the bulk extended thermodynamics for the charged AdS black hole, this topological charge $\omega=1$ means that the second-order phase transitions take place within the extended thermodynamics, mixed thermodynamics, and CFT thermodynamics, and the first-order phase transitions can emerge from these critical points. In fact, in the canonical ensemble, we can choose another characteristic function $\mathcal{F}(z)=1/\partial_{z}T(z, Q)$, which also encodes the critical points of the phase transition as pole points. The same topological charges $\omega_E=\omega_M=\omega_C=1$ can be attained. In the grand canonical ensemble, according to the principle, a characteristic function that one can choose is 
\begin{equation}\label{cffgx}
\mathcal{F}(z)=1/G_X (z),
\end{equation}
as the first-order Hawking-Page phase transition points in the bulk and the (de)confinement phase transition points on the boundary are pole points of the function. Alternatively, we can choose a characteristic function \eqref{chf} in the same form as the one chosen for the canonical ensemble, but in terms of the (bulk and boundary) electric potentials instead of the electric charges because we find that the phase transition points in the grand canonical ensemble are pole points of the function \eqref{chf}. Both the two chosen functions yield the same topological charges $\omega_E=\omega_M=\omega_C=-1$ for the phase transition points. For black holes in the modified gravity, such as the Gauss-Bonnet black holes, a more suitable characteristic function is \eqref{cffgx} to ensure that the topological charge of the Hawking-Page(-like) phase transition remains $-1$, and the differences in the topological charges of the criticalities, first-order phase transitions in the canonical ensemble, and the Hawking-Page(-like) phase transitions in the grand canonical ensemble are preserved.

Second, we can observe that in the canonical ensemble, a topological charge was assigned to a certain critical point and the same topological charge was assigned to all points on a coexistence curve of the first-order phase transition. However, in the grand canonical ensemble, one may wonder whether only a "sample point" on the one-dimensional coexistence line of the first-order Hawking-Page phase transition in the bulk or the (de)confinement phase transition on the CFT was tested. We have an answer to this question. In the bulk extended thermodynamics, for instance, one thing to note when calculating the topological charge at the critical point is that the characteristic function is given by
\begin{equation}
\mathcal{F}=\frac{8 \sqrt{\pi } \sqrt{G} z^{7/2}}{z-6 \pi  G Q^2},
\end{equation}
where $\Lambda$ is eliminated by the condition $\partial_z T(z)=0$. This procedure is similar to the one taken in \cite{Wei:2021vdx}. On the other hand, for the topology of the Hawking-Page phase transition in the bulk extended thermodynamics, for instance, the phase transition point is given by \eqref{gcese}, which corresponds to a one-dimensional coexistence line
\begin{equation}
T_c=\frac{1}{\pi}\sqrt{\frac{\Lambda}{3}\left(\Phi^2-1\right)}\label{hpco},
\end{equation}
where $T_c$ is the phase transition temperature. We notice that in the grand canonical ensemble, the characteristic function
\begin{equation}
\mathcal{F}=\frac{1}{\partial_{zz}T(z, \Phi)}=\frac{16 \pi ^{3/2} z^2 \sqrt{G z}}{G \Lambda  z-3 \pi  \left(\Phi ^2-1\right)}
\end{equation}
defined in the same way as \eqref{chf} has a pole point that happens to be the zero point of the Gibbs free energy \eqref{ginbu}. So this pole point is precisely the Hawking-Page phase transition point. Substituting this pole point into the bulk temperature in terms of $\Phi$, we obtain the coexistence line \eqref{hpco}. Therefore, all points (with different $\Phi$) on the coexistence line were tested.

\acknowledgments
We would like to acknowledge the referee for insightful comments on the paper which have been incorporated in the final version. MZ is supported by the National Natural Science Foundation of China with Grant No. 12005080.  JJ is supported by the National Natural Science Foundation of China with Grant No. 210510101, the Guangdong Basic and Applied Research Foundation with Grant No. 217200003, and the Talents Introduction Foundation of Beijing Normal University with Grant No. 310432102.

\bibliographystyle{jhep}
\bibliography{refs}
\end{document}